\documentclass[a4paper,useAMS,usenatbib]{mnras}

\usepackage{graphicx}
\usepackage{setspace}
\usepackage{natbib}
\usepackage{color}
\usepackage{amsmath,amssymb}
\usepackage{times}
\usepackage{hyperref}

\title[Unmixing the Galactic Halo with RR Lyrae tagging]{Unmixing the
  Galactic Halo with RR Lyrae tagging}

\author[Vasily A. Belokurov et al]{V. Belokurov$^{1,2}$\thanks{E-mail:vasily@ast.cam.ac.uk}, A.J. Deason$^{3}$,  S.E. Koposov$^{1,4}$, M. Catelan$^{5,6}$, D. Erkal$^{1,7}$, A.J. Drake$^8$,\newauthor and N.W. Evans$^1$\\
  $^{1}$Institute of Astronomy, Madingley Rd, Cambridge, CB3 0HA\\
  $^{2}$Center for Computational Astrophysics, Flatiron Institute, 162 5th Avenue, New York, NY 10010, USA\\
  $^{3}$Institute for Computational Cosmology, Department of Physics, University of Durham, South Road, Durham DH1 3LE, UK\\
  $^4$Department of Physics, McWilliams Center for Cosmology, Carnegie Mellon University, 5000 Forbes Avenue, Pittsburgh, PA 15213, USA\\
  $^{5}$Instituto Milenio de Astrofísica, Santiago, Chile\\
  $^{6}$Departamento de Astronom\'ia, Casilla 160-C, Universidad de Concepci\'on, Casilla 160-C, Concepci\'on, Chile\\
  $^{7}$Department of Physics, University of Surrey, Guildford GU2 7XH, UK\\
  $^8$California Institute of Technology, 1200 E. California Blvd, CA 91225, US}

\begin{document}


\maketitle

\label{firstpage}

\begin{abstract}
We show that tagging RR Lyrae stars according to their location in the
period-amplitude diagram can be used to shed light on the genesis of
the Galactic stellar halo. The mixture of RR Lyrae of ab type,
separated into classes along the lines suggested by Oosterhoff,
displays a strong and coherent evolution with Galactocentric
radius. The change in the RR Lyrae composition appears to coincide
with the break in the halo's radial density profile at $\sim$25
kpc. Using simple models of the stellar halo, we establish that at
least three different types of accretion events are necessary to
explain the observed RRab behavior. Given that there exists a
correlation between the RRab class fraction and the total stellar
content of a dwarf satellite, we hypothesize that the field halo RRab
composition is controlled by the mass of the progenitor contributing
the bulk of the stellar debris at the given radius. This idea is
tested against a suite of cosmological zoom-in simulations of Milky
Way-like stellar halo formation. Finally, we study some of the most
prominent stellar streams in the Milky Way halo and demonstrate that
their RRab class fractions follow the trends established previously.

\end{abstract}


\section{Introduction}

``Chemical tagging'' postulates the existence of a stellar fingerprint
--- a unique pattern of elemental abundances --- that each star carries
and that can be used to trace it back to its location of origin
\citep[][]{Freeman2002}. This hypothesis has motivated many
spectroscopic surveys
\citep[][]{Gilmore2012,Allende2008,Majewski2017,Desilva2015, LAMOST}
and has kick-started a number of observational
\citep[e.g.][]{Helmi2006,Majewski2012,Blanco2015} and theoretical
\citep[e.g.][]{Font2006,Roskar2008,Bland2010} investigations, opening
a new field of astrophysical enquiry --- Galactic chemo-dynamics
\citep[see e.g.][]{Ralf2009,Minchev2013, Minchev2014}. The application
of the ``chemical tagging'' idea to the studies of the Galactic disc
has enjoyed plenty of success
\citep[e.g.][]{Desilva2007,Bensby2014,Bovy2016}, yet inevitably, has
uncovered a number of stumbling blocks too \citep[see
  e.g.][]{Mitschang2014, Ting2015, Ness2017}.

In the Milky Way's halo, progress has so far been much slower, mainly
because acquiring a large number of high resolution spectra of stars
across a range of halo locations remains prohibitively arduous. For
example, the most recent attempt to survey the halo at high resolution
beyond the Solar neighborhood \citep[see][]{Battaglia2017} includes
only 28 stars, whose heliocentric distances are mostly limited to
within 30 kpc. Note, however, that while the field halo remains poorly
explored (in terms of precision chemistry), there exists a large
amount of spectroscopic data on the surviving satellites. Curiously,
while similar in concept, the genealogy of the disc and the halo
differ substantially in practice. In the disc, chemo-dynamicists
attempt to rewind the stars back to the low-mass star clusters they
were born in. In the halo, a much richer variety of progenitors are
available: stellar systems with masses from that of a giant molecular
cloud to that of the Large Magellanic Cloud (LMC) could have all
contributed to the halo formation.

For example, by comparing the abundance trends in the globular clusters
(GCs) and the local halo stars, several authors claim that as much as
$50\%$ of the stellar halo could have originated in GCs
\citep[][]{Caretta2010,Martell2010,Martell2011}. The above calculation
relies heavily on the theory that the GCs were 10-20 times more
massive in their youth and have experienced prolific mass loss since ---
the argument put in place to explain the ratios between the second and
first generations of their member stars \citep[see][for a
  review]{Gratton2012}. However, the study of \citet{Deason2015} argues
against the stellar halo creation through GC disruption. They base
their argument on the measurements of the ratio of the number of Blue
Horizontal Branch stars to that of Blue Stragglers. Bear in mind
though that, as \citet{Chung2016} show, this constraint may perhaps be
circumvented by invoking the loss of most of the first generation
stars early in a cluster's life.

At the other end of the mass spectrum, observational evidence is
mounting for the most massive dwarf galaxies to contribute a
significant fraction of the stellar halo. First, according to the
$\alpha$-elements to iron abundance ratio (as gleaned both from low
and high resolution spectroscopy), the local stellar halo appears
similar to galaxies like the LMC and Sgr \citep[see
  e.g.][]{Venn2004,Tolstoy2009,Deboer2014}. Additionally, the Milky
Way (MW) halo's radial density profile shows a break at around 25 kpc
\citep[][]{Watkins2009,Deason2011,Sesar2011}, which according to
\citet{Deason2013} could be best explained with an early accretion of
a massive stellar system. The final clue can be found in the study of
the make-up of the RR Lyrae population of the Galactic halo.  For
example, \citet{Fiorentino2015} demonstrate that high amplitude short
period (HASP) RRab stars can be used to decipher the relative
contributions of the GCs, Ultra-Faint dwarfs (UFDs), classical dwarf
spheroidals (dSph) and massive systems such as the LMC and the SMC. They
point out that while the HASP RR Lyrae are a common denizen of the
halo, these stars are completely lacking in most surviving dwarfs
(including all UFDs), except for the Sagittarius (Sgr) and the Magellanic
Clouds. In GCs, as \citet{Fiorentino2015} show, only the metal-rich
systems have sizeable HASP populations. Thus the only pathway to
HASP creation is via a massive system with rapid metal
enrichment. This picture is in full agreement with the earlier
analysis of the period-amplitude distribution (known as the Bailey
diagram) of the RRLs in the halo and the MW satellites \citep[see
  e.g.][]{Catelan2009,Zinn2014}.

Note that the hypothesis of the stellar halo creation by way of a
massive dwarf galaxy disruption does not directly contradict the
evidence for a substantial GC
contribution. Incontrovertibly, GCs do not form on their
own but always require a host, an unlucky galaxy destined to be
tidally destroyed \citep[see
  e.g.][]{Krui2015,Bekki2016,Boylan2017,Renaud2017}. Note that these
models also allow formation of some of the metal-rich clusters in-situ
\citep[also see formation scenarios discussed
  in][]{Caretta2010}. However, it seems likely that an in-situ stellar
halo would possess some residual spin still observable at the present
day. Indeed, in the MW, there have been some claims of a
detection of an in-situ halo population \citep[see e.g][]{Carollo2007,
  Carollo2010}. However, given that no substantial rotation has been
reported for the Galactic metal-poor halo tracers
\citep[see][]{Deason2017spin}, it is safe to assume that the in-situ
contribution to the MW old halo at high Galactic $|z|$ is minimal.

Importantly, given that the debris mixing times are a strong function
of Galactocentric radius and that the dynamical friction depends
strongly on the satellite's mass and its orbital parameters, it is
na\"ive to expect the properties of the stellar halo to be the same
across the Galaxy. Unfortunately, the exploration of the evolution in
the stellar halo's make-up has been hindered by the excessive cost of
running a spectroscopic survey of such an enormous volume at such a
low target density. In this Paper, we propose to decipher the
fractional contributions of stellar systems of different masses by
mapping out the change in the mixture of pulsating RR Lyrae variables.

Our proposed tagging scheme relies on the ideas of Peter Oosterhoff
\citep[see][]{Oo1939,Oo1944} who pointed out striking differences in
the period distributions of the RRab stars in GCs. Clusters appear to
be separable into two classes, one with a mean period of $\sim$0.55
days and another with a period of $\sim$0.65 days. The exact
explanation of this so-called ``Oosterhoff dichotomy'' remains to be
found \citep[but
  see][]{Stell1975,Lee1990,Bono1997,Clement1999,Jurcsik2003,Gratton2010,
  Sollima2014}. It has been demonstrated, however, that the stars in
the two groups show small but significant differences in pulsation
amplitude, color and metallicity, which are related to differences in
their masses, luminosities, effective temperatures and iron and helium
abundances \citep[see][]{vanAlbada1973,Sandage2004}. Based on the
holistic examination of the properties of GCs falling into two
Oosterhoff groups, the consensus appears to be that the period
dichotomy in clusters reflects the difference in their age ---
e.g. Oosterhoff I objects appear on average younger compared to those
of Oo II type \citep[see][]{vdb1993,vdb2011} --- and metallicity
\citep[e.g.][]{Sandage1993} and thus, most likely, differences in
their birth environment \citep[see e.g.][]{Lee1999}. Lending further
support to the hypothesis of a possible correlation between the
Oosterhoff type and the object's origin is the lack of RR Lyrae
belonging to either of the classes in the surviving dSph satellites of
the MW. Although it has recently been suggested that internal helium
abundance variations brought about by the presence of multiple
populations can also play a role in explaining the phenomenon
\citep[see e.g.][and references therein]{Jang2015, Vandenberg2016}. As
displayed in e.g. \citet{Catelan2009}, \citep[see also][for a more
  recent discussion]{Catelan2015}, dwarfs appear to be mostly
Oosterhoff-intermediate, i.e. falling in-between the two classes as
gleaned from the GC analysis.

Presented with the distinct behaviour of RR Lyrae in the
period-amplitude space as described above, we put forward a simple
observational diagnostic for the genesis of the Galactic field halo
population. We propose to shed light onto the likely halo progenitors
by mapping out the fraction of the pulsators occupying different
locations in the Bailey diagram. This analysis is contingent on the
fact that the MW satellites do not contain RRab stars of a particular
Oosterhoff type, but rather host mixtures of these
\citep[][]{Catelan2009}. Here, we concentrate in particular on the
difference in the 3D distribution of the RRab stars that fall
approximately into the Oosterhoff I and II groups (see
Section~\ref{sec:rrldata} for details). Moreover, motivated by the
results presented in \citet{Fiorentino2015}, we compare the behavior
of these two classes with the spatial evolution of the fraction of the
HASP variables. Note that compared to the multitude of the in-depth
studies of the Oosterhoff dichotomy in the surviving satellites of the
Galaxy, little has been done so far with regards to the field RR Lyrae
population. This is not an omission but a delay due to the lack (until
recently) of large all-sky RR Lyrae samples.

\begin{figure}
  \centering
  \includegraphics[width=0.48\textwidth]{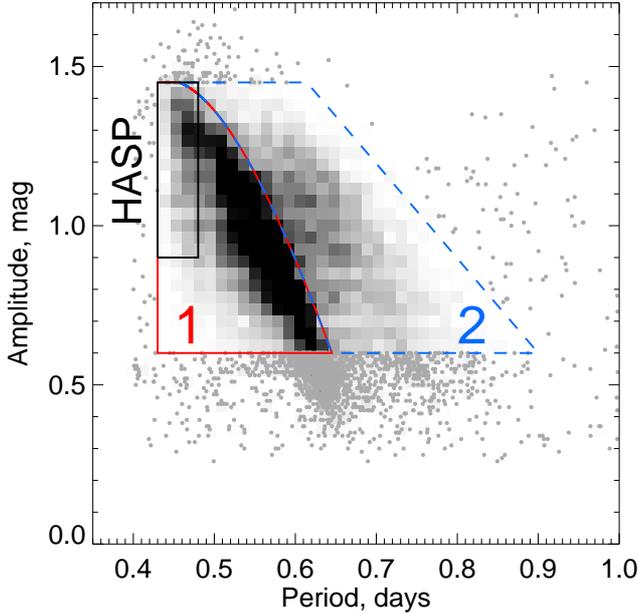}
  \caption[]{Density of CRTS RR Lyrae stars in the plane of V-band
    amplitude (in magnitudes) and period (in days). Locations of the
    Type 1 (red), Type 2 (blue) and the High Amplitude Short Period
    (HASP, black) objects are shown. Selection boundaries are
    stipulated in the main text. Note that according to this
    selection, Type 1 objects are predominantly Oosterhoff I RR Lyrae,
    while Type 2 is composed of Oosterhoff II and Oosterhoff
    intermediate objects.}
   \label{fig:select}
\end{figure}

While precious few in number, several previous studies of the
period-amplitude make-up of the MW field population exist. For
example, the Oosterhoff dichotomy in the Galactic halo RRab stars is
studied in \citet{Miceli2008}, who find that within 20-30 kpc from the
Galactic center, the Oosterhoff II RRabs follow a steeper radial
density law compared to those belonging to the Oosterhoff I
group. \citet{Zinn2014} explore not only the field halo but also the
prominent halo sub-structures crossing the field-of-view of their
$\sim800$ deg$^2$ survey, such as the Sgr stream and the Virgo
Stellar Stream (VSS). They measure the shape of the stellar halo and
see a clear break (see the discussion above) in the radial density law
at $\sim$25 kpc. While \citet{Zinn2014} detect no noticeable change in
the Oosterhoff mixture across the break, they caution that this could
simply be due to the small RRL sample size available to
them. Concentrating on the Sgr stream, they notice that the stream
contains a smaller fraction of short period pulsators (compared to the
remnant) and link this to the chemical abundance gradients in the
progenitor. Most importantly, they point out a great level of
similarity between the period-amplitude distribution in the field halo
and in the large sub-structures such the Sgr stream and the
VSS. Moving closer to the center of the Galaxy, \citet{Kunder2009}
scrutinize the bulge RR Lyrae population and notice an apparent
difference between the period-amplitude distribution of the bulge RR
Lyrae and of those elsewhere in the MW. More precisely, they
observe a much higher fraction of shorter period objects (at fixed
amplitude), which they link to an enhanced metal enrichment,
\citep[also see][for an updated
  analysis]{Kunder2013,Pietru2015}. Based on the observed difference,
\citet{Kunder2009} conjecture that the progenitor of the Galactic
bulge ought to be distinct from the halo parent system(s).

This Paper is structured as follows. In Section~\ref{sec:rrldata}, we
describe the sources of the RR Lyrae data used here as well as the
selection boundaries in the period-amplitude space. We show how the
mixture of RRab stars from different portions of the Bailey diagram
evolves with Galactocentric radius in Section~\ref{sec:radial}. The
Oosterhoff dichotomy in the currently known most prominent stellar
streams is discussed in Section~\ref{sec:streams}. We also look at the
amount of small-scale clustering of RRab stars in the halo as a
function of their period-luminosity location in
Section~\ref{sec:cluster}. Finally, we provide the summary and the
context for this study in Section~\ref{sec:disc}.

\section{RR Lyrae data}
\label{sec:rrldata}

This works relies, in part, on the all-sky RR Lyrae data recently made
publicly available by the Catalina Rapid Transient Survey (CRTS). The
CRTS sample was released in several installments; we use the two
largest subsets, namely the Northern sky as reported in
\citet{Drake2013} and the complementary Southern portion analysed in
\citet{Torrealba2015}. After the cross-match of the two catalogs,
22,700 unique RR Lyrae survive. The distribution of their periods and
amplitudes is shown in Figure~\ref{fig:select}. Note that in this
analysis, the CRTS amplitudes are corrected by 0.15 mag as explained
in \citet{Drake2013}. Here, we use the RR Lyrae distances as published
in the above catalogs. To convert to the Galactocentric, we assume
a solar radius of 8 kpc. Only RRab stars are considered in this
work. These are selected according to the following Period (P) and
Amplitude (Amp) criteria:

\begin{equation}
  \label{eq:select1}
  \begin{array}{r@{}l}
  0.43 < {\rm P} < 0.9\\
  0.6 < {\rm Amp} < 1.45\\
  {\rm Amp} < 3.75-3.5{\rm P}
  \end{array}
\end{equation}

\noindent This first cut ensures that, compared to the bulk of the
RRab stars, possible contaminating variables with shorter (a minuscule
contamination from RRc stars at short periods and small amplitude is
still possible) or longer periods are excluded. Note that we have
checked that the results presented below do not change significantly
if the lower amplitude boundary is increased. The second cut attempts
to minimise the effects of the CRTS selection efficiency related to
the pulsation amplitude. Only stars with Amp$>$0.6 are detected out to
50 kpc without a significant loss in completeness. There are $N_{\rm
  tot}=20,839$ RRab stars available after this selection. The
Oosterhoff Type I and Type II stars are demarcated by the line given
in Equation~11 of \citet{zoro2010} but offset by +0.15 in amplitude,
namely

\begin{equation}
  \label{eq:select2}
  {\rm Amp} = -2.477-22.046\log{\rm P}-30.876(\log {\rm P})^2
\end{equation}

\noindent As evidenced by Figure~\ref{fig:select}, this selection
boundary wraps tightly around the over-density of Oosterhoff~I
stars. Note that according to the above definition, the Class 2 will
include both Oosterhoff II and Intermediate objects. From here
onwards, we refer to the RRab stars falling within these selection
boxes as Type (or Class) 1 and 2 stars.

Finally, we select High Amplitude Short Period (HASP) RR Lyrae
according to the following conditions:

\begin{equation}
  \label{eq:select3}
  \begin{array}{r@{}l}
  0.43 < {\rm P} < 0.48\\
  0.9 < {\rm Amp} < 1.45\\
  \end{array}
\end{equation}

We complement the CRTS data with a sample of RR Lyrae stars in
globular cluster (GC) and dwarf spheroidal (dSph) satellites of the MW. The GC RR Lyrae are taken from the updated version
\citep[see][]{Clement2017} of the catalog presented by
\citet{Clement2001}. Finally, we also use the following catalogues of
RR Lyrae stars in the MW dSphs: in Cetus and Tucana by
\citet{Bernard2009}, in Draco by \citet{Kinemuchi2008}, in Fornax by
\citet{Greco2009}, in Leo 1 by \citet{Stetson2014}, in Sculptor by
\citet{Kaluzny1995}, in the LMC by \citet{Igor2003} and in the SMC by
\citet{Kapakos2011}. For the Sgr dwarf, we use the catalogue of
\citet{Igor2014} and extract all RR Lyrae within 10 degrees from the
center of the dwarf and located between 22 and 31 kpc from the Sun.

\section{RRab mixture evolution with Galactocentric radius}
\label{sec:radial}

\begin{figure}
  \centering
  \includegraphics[width=0.48\textwidth]{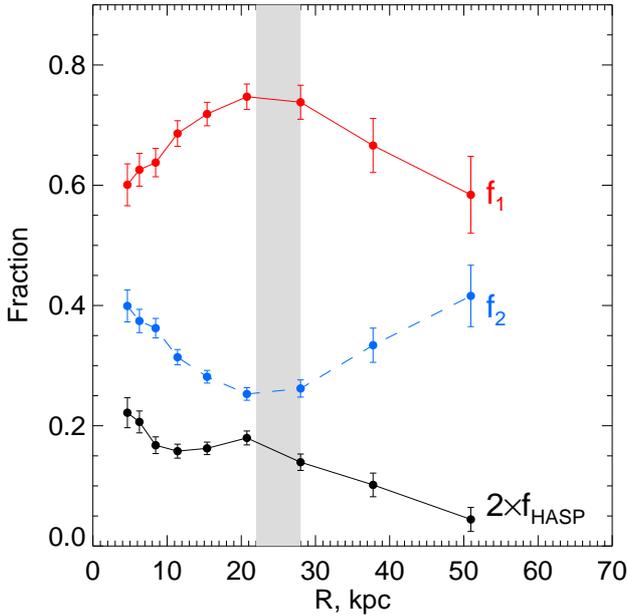}
  \caption[]{The RRab mixture evolution as a function of
    Galactocentric radius (kpc). Fractions $f_1$ (red), $f_2$ (blue)
    and $f_{\rm HASP}$ (black, scaled up by a factor of 2) are each
    shown to vary significantly with radius. Note that $f_{\rm HASP}
    +f_1+f_2 \ne 1$ as HASP RRab are (almost entirely) a subset of the
    Type 1 objects. The red and blue curves are mirror images of each
    other by definition, whereas the HASP fraction $f_{\rm HASP}$
    behaves differently from both $f_1$ and $f_2$. The thick grey
    vertical line marks the location of a break in the stellar halo
    radial density profile
    \citep[see][]{Watkins2009,Deason2011,Sesar2011}. The stars
    belonging to the Sgr stream (see Section~\ref{sec:streams}) have
    been excluded by excising all objects with stream latitude
    $|B|<8^{\circ}$, where the stream coordinates are obtained by
    rotating RA, Dec to align with the great circle with a pole at
    $(\alpha, \delta)=(303.63^\circ, 59.68^\circ)$.}
   \label{fig:profile}
\end{figure}
\begin{figure}
  \centering
  \includegraphics[width=0.48\textwidth]{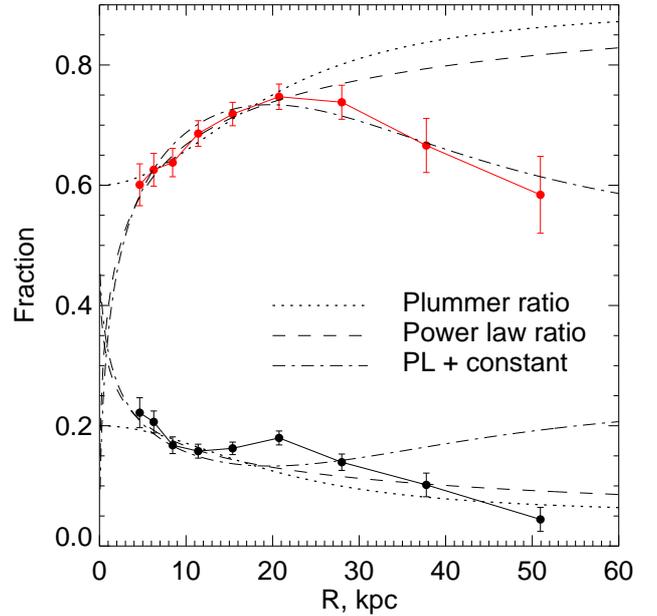}
  \caption[]{$f_1$ and $f_{\rm HASP}$ radial profiles (as shown in
    Figure~\ref{fig:profile}) with toy mixture models overlaid. The dashed
    line shows the $f_1$ fraction in the halo where Type 1 and Type 2
    RR Lyrae are distributed according to (spherical) power-law models
    reported in \citet{Miceli2008}, namely: power law index -2.7 for
    Type 1 and -3.2 for Type 2. The dash-dotted line corresponds to a
    power-law density model as above but with an additional constant
    density component. The dotted line is a halo where each RRab type is
    represented with a Plummer density model: Type 1 with a scale
    radius of 30 kpc, and Type 2 with 21 kpc.}
   \label{fig:profile2}
\end{figure}

Figure~\ref{fig:profile} displays the change in the fraction of Type 1
(red) RR Lyrae with respect to all stars selected using
Equations~\ref{eq:select1}, e.g. $f_1(R)=N_1(R)/N_{\rm tot}(R)$, as a
function of Galactocentric radius. By design, $N_{\rm tot}=N_1+N_2$,
thus $f_2=1-f_1$. All type fraction come with their associated
uncertainties, which are computed by propagating the Poisson errors.
Note that for this and other Figures showing the change in the RR
Lyrae composition with radius, we apply additional cuts in Galactic
height $z$ and extinction:

\begin{equation}
  \label{eq:select4}
  \begin{array}{r@{}l}
  E(B-V) < 0.25\\
  |z| > 1~{\rm kpc}\\
  \end{array}
\end{equation}

\noindent We also exclude the most significant halo sub-structure,
i.e. the Sgr Stream (see Figure caption for details). After the three
extra cuts, we are left with $N_{\rm tot}=15,454$ RRab stars, of these
$10,676$ are Type 1, $4778$ are Type 2 and $1297$ are HASP. Also shown in
Figure~\ref{fig:profile} is the fraction of Type 2 (HASP) objects in
blue (black). While the CRTS completeness is a strong function of
magnitude and to lesser extent of position on the sky, we believe that
the selection biases affect in equal measure the stars in the three
groups considered. This, of course, assumes that there are no dramatic
differences in the amplitude distributions of the two
classes. However, based on our experiments with the sample selection
boundaries, we believe that the fraction curves displayed indicate the
actual change in the RR Lyrae mixture throughout the Galaxy. Across
the distance range allowed by the data (5-50 kpc), the Type 1 objects
dominate, comprising approximately two thirds of the overall RR Lyrae
population, in accordance with previous studies
\citep[see][]{Miceli2008, Abbas2014, Zinn2014}. As the Figure
evidently demonstrates, this fraction also varies significantly with
radius: $f_1$ drops near the Galactic center, and beyond 30 kpc. The
peak in $f_1$ fraction, somewhere between 20 and 30 kpc, appears to
match the location of the break in the stellar halo's radial density
profile \citep[see][]{Watkins2009, Sesar2011, Deason2011}.

\begin{figure*}
  \centering
  \includegraphics[width=0.32\textwidth]{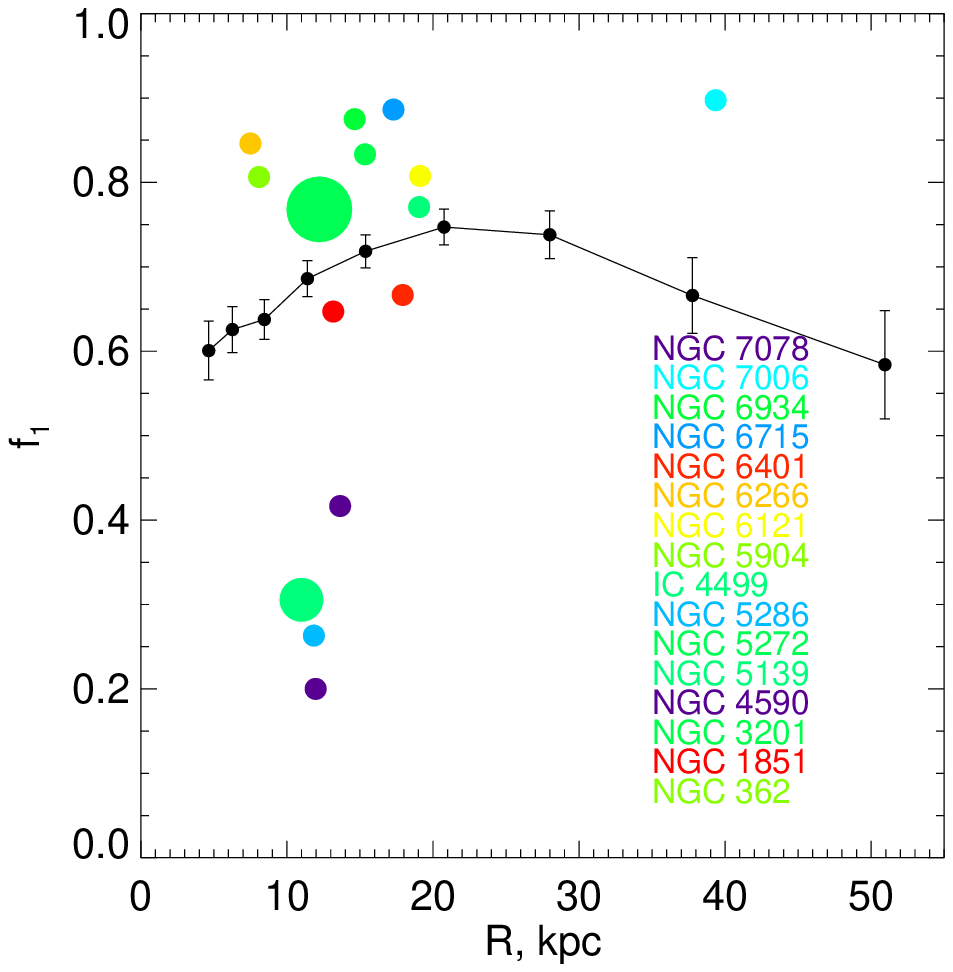}
  \includegraphics[width=0.32\textwidth]{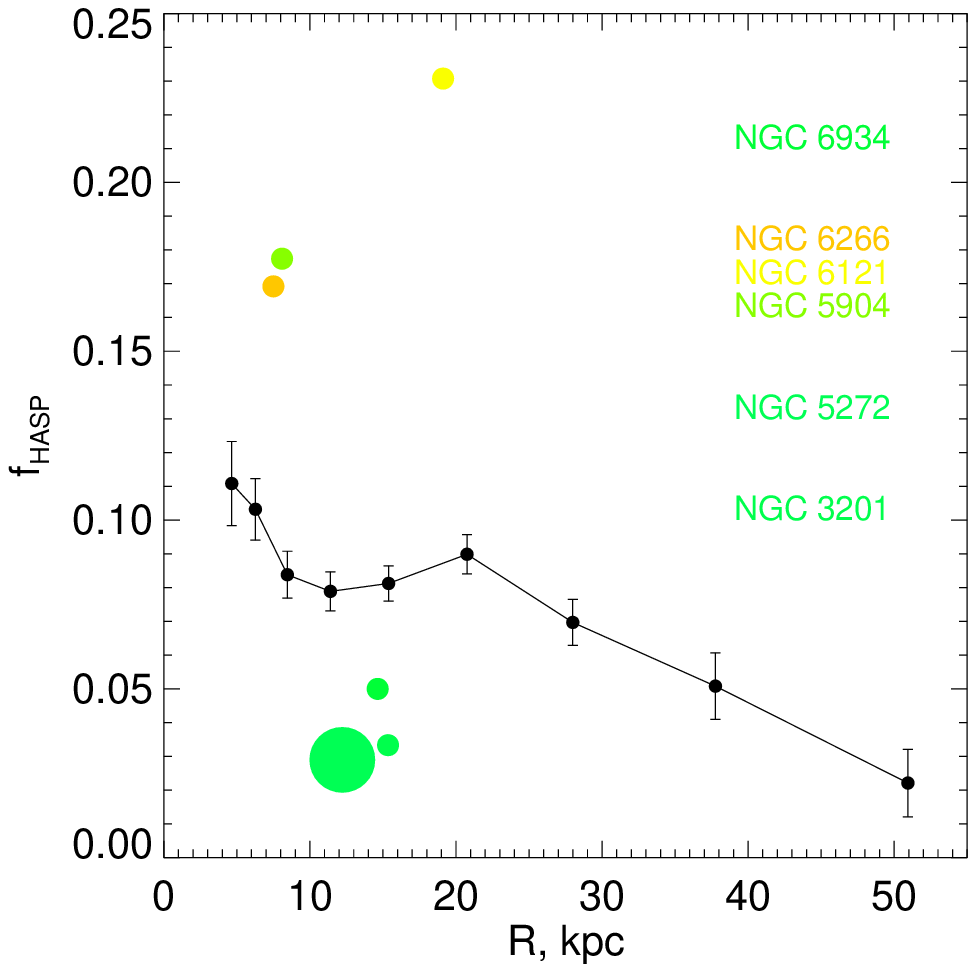}
  \includegraphics[width=0.32\textwidth]{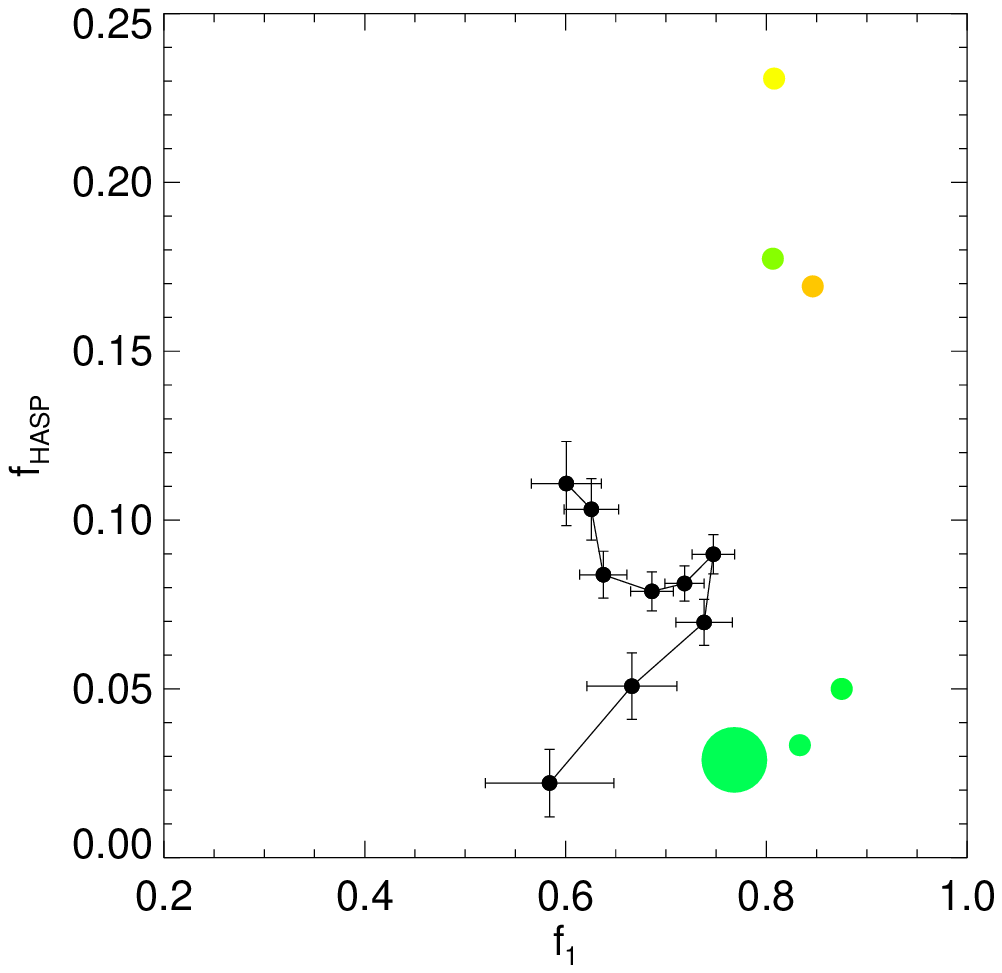}
  \caption[]{{\it Left:} Fraction of Class 1 RRab stars in the Milky
    Way Globular Clusters. The size of the circle represents the
    number of RR Lyrae in the cluster, while the color encodes the
    GC's metallicity, with violet/blue being the most metal-poor and
    orange/red the most metal-rich. Note the well-known Oosterhoff
    dichotomy where GCs tend to avoid intermediate values of
    $0.4<f_1<0.7$. {\it Middle:} Fraction of High Amplitude Short
    Period RRab stars $f_{\rm HASP}$ in GCs. Very few, typically
    metal-rich, objects achieve $f_{\rm HASP}>0.15$. Note that only
    objects with more than 1 HASP star are included in this
    panel. {\it Right:} The distribution of GCs in the space spanned
    by $f_1$ and $f_{\rm HASP}$. In all three panels, the black solid line
    shows the measurement of the Galactic stellar halo.}
   \label{fig:gcs}
\end{figure*}
\begin{figure*}
  \centering
  \includegraphics[width=0.32\textwidth]{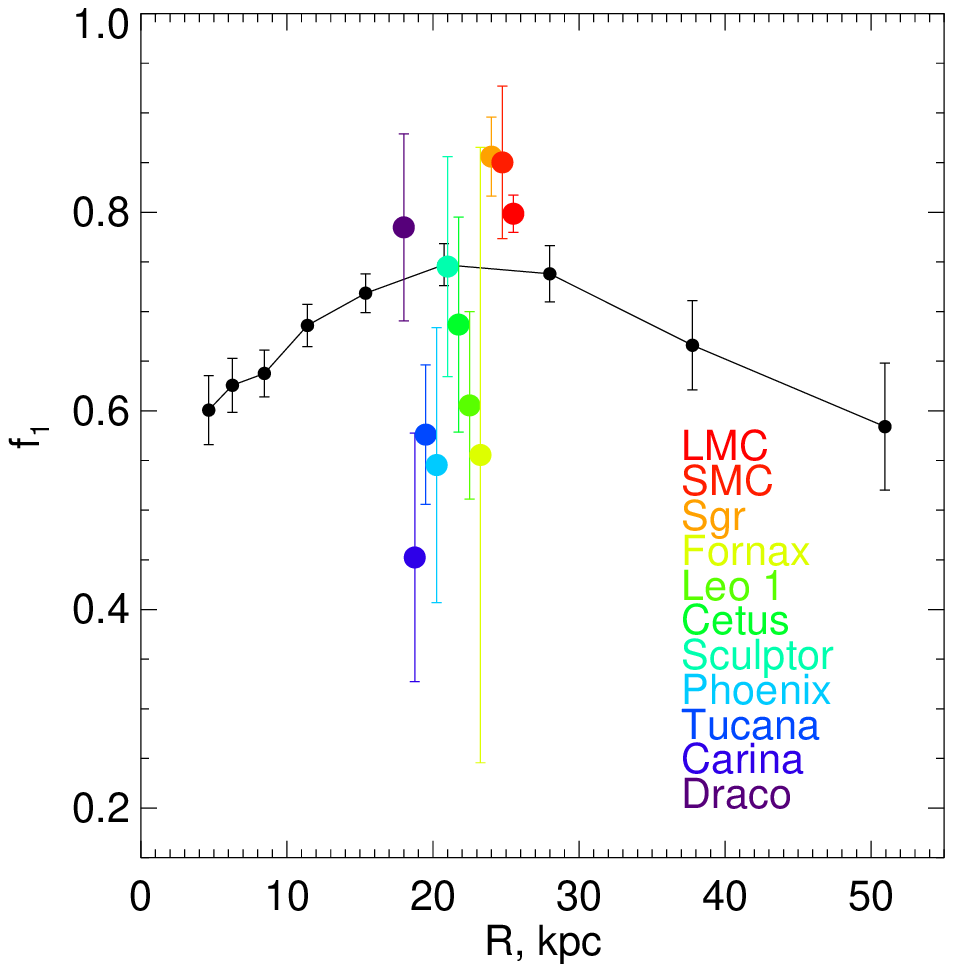}
  \includegraphics[width=0.32\textwidth]{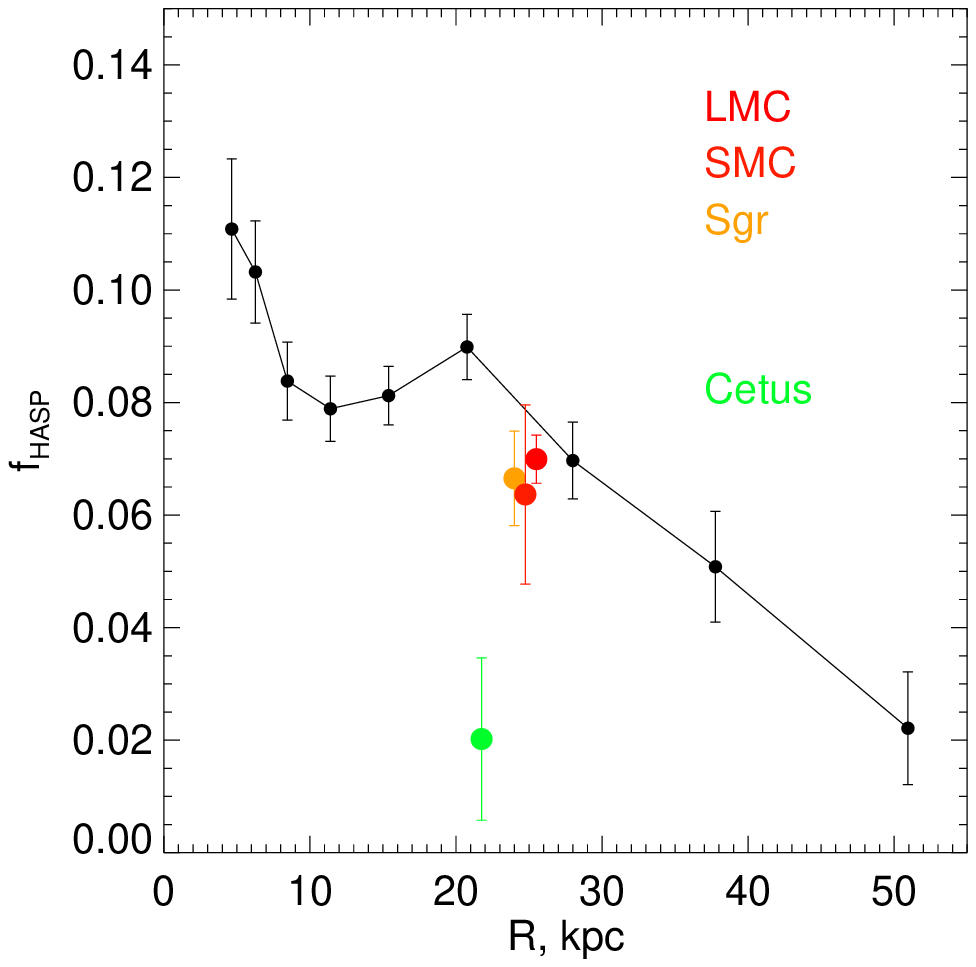}
  \includegraphics[width=0.32\textwidth]{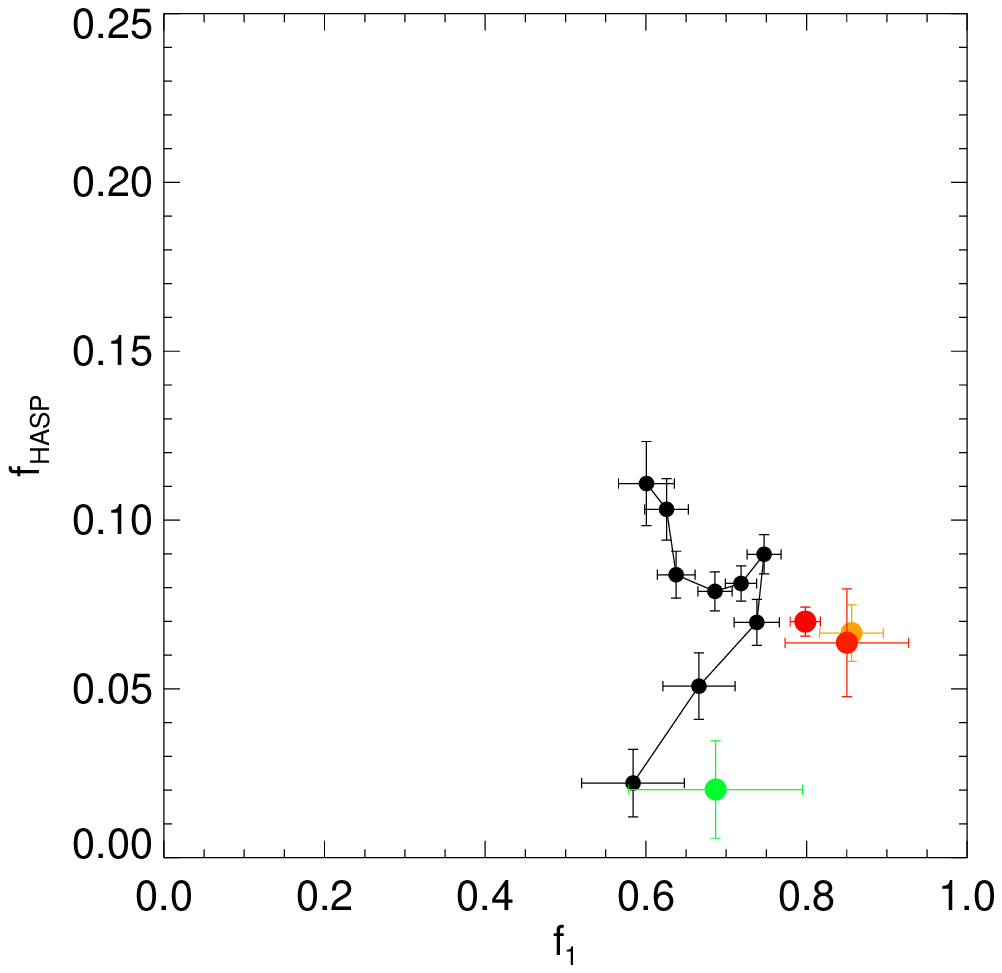}
  \caption[]{Same as Figure~\ref{fig:gcs} for surviving Galactic
    dwarf spheroidals. Note that the color coding is different here,
    with color representing the satellite's stellar
    mass. Additionally, the dSphs are not shown at their current
    distances in the Galaxy, but are instead offset to the fiducial
    $R\sim20$ kpc for comparison with the stellar halo measurement
    (solid black). Note a correlation between $f_1$ and dwarf's
    stellar mass as shown in the Left panel. While the surviving
    dwarfs --- occupying the so-called Oosterhoff intermediate regime ---
    match well the range of the stellar halo's $f_1$, none of the
    currently observable systems reach the high levels of $f_{\rm
      HASP}$ reported for the field halo.}
   \label{fig:dsph}
\end{figure*}

Evidently more dramatic is the change in the HASP fraction. The
proportion of the HASP RR Lyrae changes by a factor 5 from 10$\%$
around the Galactic center to 2$\%$ in the outer halo. Intriguingly,
the $f_{\rm HASP}$ behavior does not match fully the evolution of
$f_1$: while, similar to Type 1 objects, the proportion of HASP RR
Lyrae peaks around 25 kpc, there is an even stronger increase towards
low $R$, in the range where $f_1$ keeps declining.

To illustrate the differences in the behavior of Type 1/Type 2/HASP RR
Lyrae with radius, Figure~\ref{fig:profile2} presents simple toy
models of the fraction evolution: a ratio of power-law (Plummer)
density profiles in dashed (dotted) black curves. For the power-law
density, we use the indices measured by \citet{Miceli2008}, and given
in their Equations 28-29, namely, $-2.7$ for Class 1 and $-3.2$ for Class
2. The density profiles suggested by \citet{Miceli2008} appear to give
a reasonable description of the $f_1$ behavior within the distance
range probed by their data. Beyond 30 kpc, however, this model does
not agree with the CRTS data: it predicts a continuing increase in the
fraction of Type 1 stars, while a clear drop is registered. Using a
ratio of two Plummer models with scale radii 30 kpc (21 kpc) for Class
1 (2) achieves a marginal improvement, in particular in the very inner
portions of the halo, from 5 to 15 kpc.

Given the $f_1$ behaviour out to 30 kpc --- as shown in
Figures~\ref{fig:profile} and \ref{fig:profile2} --- one simple
explanation of the change in the Galactic RRab mixture is to
invoke two types of progenitors. In this scenario, a stellar system
with a higher fraction of Type 1 RR Lyrae (Component 1) deposits tidal
debris which then relaxes into a distribution with a flatter density
profile as compared to the stars left behind by the progenitor(s) with
a higher fraction of Type 2 RRab stars (Component 2). Please note that
the models shown in Figure~\ref{fig:profile2} assume an extreme case
where one type of progenitor contributes all RR Lyrae of a particular
type. However, from our experiments with these simple models, it is
clear that more realistic values $f_1\sim0.2$ would also hold well
against the data. This simple picture gives a convincing explanation
of the fractional increase in Type 2 objects at small distances, but
seemingly breaks down at large Galactocentric distances, where the
$f_1$ is observed to drop again.

\begin{figure*}
  \centering
  \includegraphics[width=0.98\textwidth]{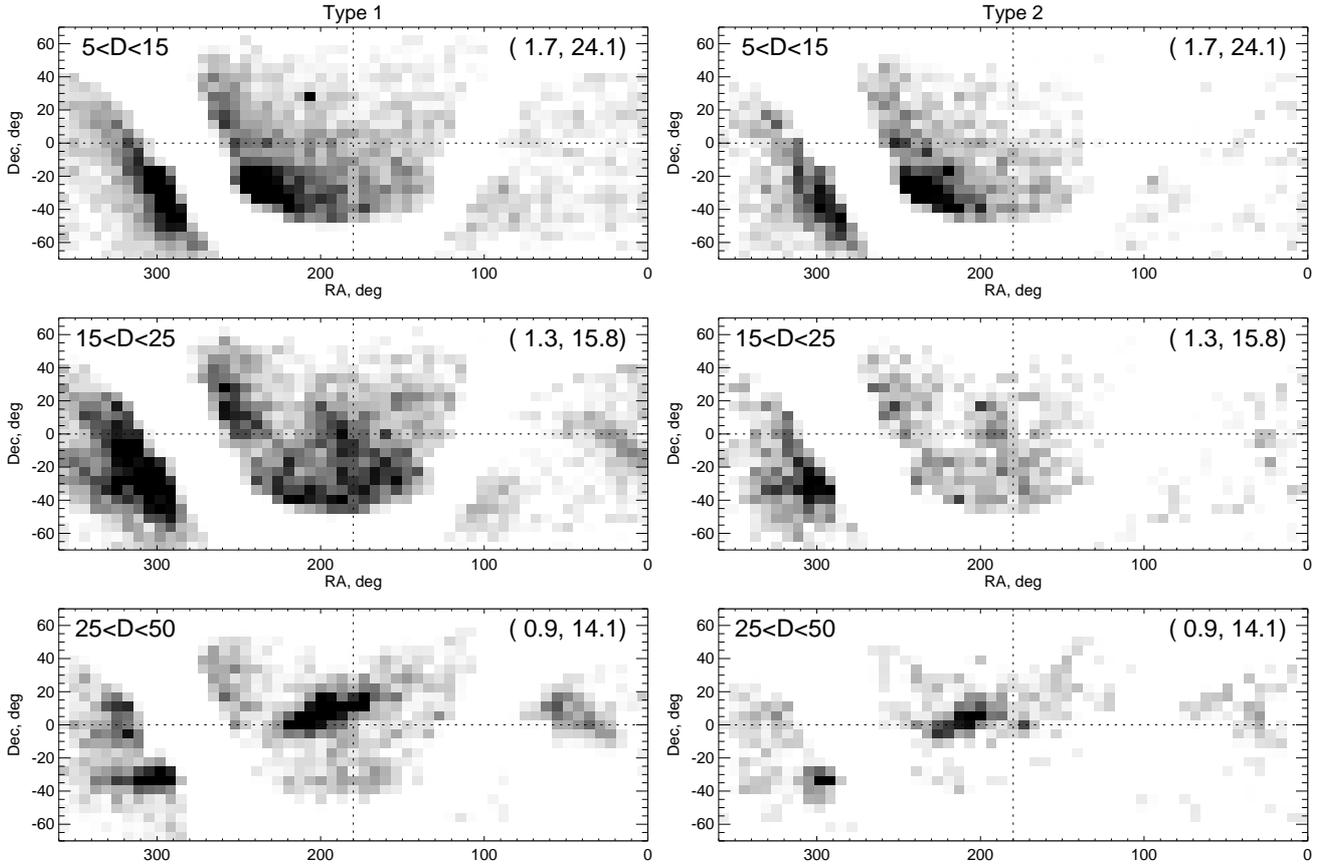}
  \caption[]{Density of RRab stars of Type 1 (left) and Type 2 (right)
    in equatorial coordinates for three heliocentric distance range
    (shown in the top left corner of each panel). There are 55 pixels
    in the RA dimension and 25 along Dec. The maps are smoothed with a
    Gaussian kernel with FWHM of 1.1 pixels. The number of stars per
    pixel corresponding to the black and white shades of grey is shown
    in the top right corner of each panel.}
   \label{fig:halomap}
\end{figure*}

Such a drop in $f_1$ fraction (or, equivalently, increase in $f_2$)
beyond the break radius could possibly be accounted for with an
addition of a third progenitor type - that with an extremely flat
radial density distribution, as illustrated by the dash-dotted line in
Figure~\ref{fig:profile2}. This model has an additional component
whose density is constant with radius. As evident in the Figure, the
contribution of the third component appears to be enough to explain
the turnover in the $f_1$ curve beyond 30 kpc. There is perhaps a
more prosaic explanation for the peak and the turnover of the $f_1$
curve. Rather than requiring an additional component, such behavior
could probably be the result of the difference in sphericity between
the Component 1 and 2 debris. The change in the stellar halo
flattening with Galactocentric radius has recently been reported in a
number of studies \citep[see][]{Xue2015,Das2016,Iorio2017}. Quite
simply, a significant vertical (with respect to the Galactic disc)
flattening of the Component 1 debris would likely result in a decrease
of $f_1$ at large radii. Finally, if, as argued by \citet{Deason2013},
the break in the stellar halo density is created by the accretion of
one massive stellar system, the Component 1 might not be extending
much beyond the apo-centre of the host satellite, thus yielding a
sharp truncation at around the break and subsequent drop in
$f_1$. This scenario, of course, works very well in conjunction with
the previous hypothesis, i.e that of a difference in the component
flattening.

The evolution of the fraction of Type 1 RRab stars can be contrasted
with the change in the fraction of HASP pulsators, shown as a black
solid line in Figures~\ref{fig:profile} and \ref{fig:profile2}. The HASP
fraction evolves in the opposite sense to $f_1$ and out to $\sim$15
kpc follows the trend in $f_2$ fraction. Beyond that, it switches from
following $f_2$ and starts to track $f_1$. Apart from a bump in the
$f_{\rm HASP}$ profile at $\sim$20 kpc, it continues to fall
precipitously as one moves into the outer halo. In summary, given this
distinct evolution, changes in $f_{\rm HASP}$ cannot be described by
the simple two-component model proposed above. Nonetheless, to
illustrate the differences and commonalities between $f_1$, $f_2$ and
$f_{\rm HASP}$, Figure~\ref{fig:profile2} shows the inverted versions
of the power law and the Plummer models described above. Clearly, both
power-law density ratio and Plummer density ratio can explain crudely
the global shape of $f_{\rm HASP}$ curve. However, the rise of the
HASP fraction at $\sim$20 kpc remains unaccounted for. Alternatively,
a three component model can match the rise at the halo break radius,
but does not have the subsequent fall-off in the outer halo. Also note
that the rise and the fall in $f_{\rm HASP}$ fraction happens much
faster compared to the evolution of $f_1$ and $f_2$.

\subsection{RRab mixture in Galactic satellites}
\label{sec:sats}

\begin{figure*}
  \centering
  \includegraphics[width=0.33\textwidth]{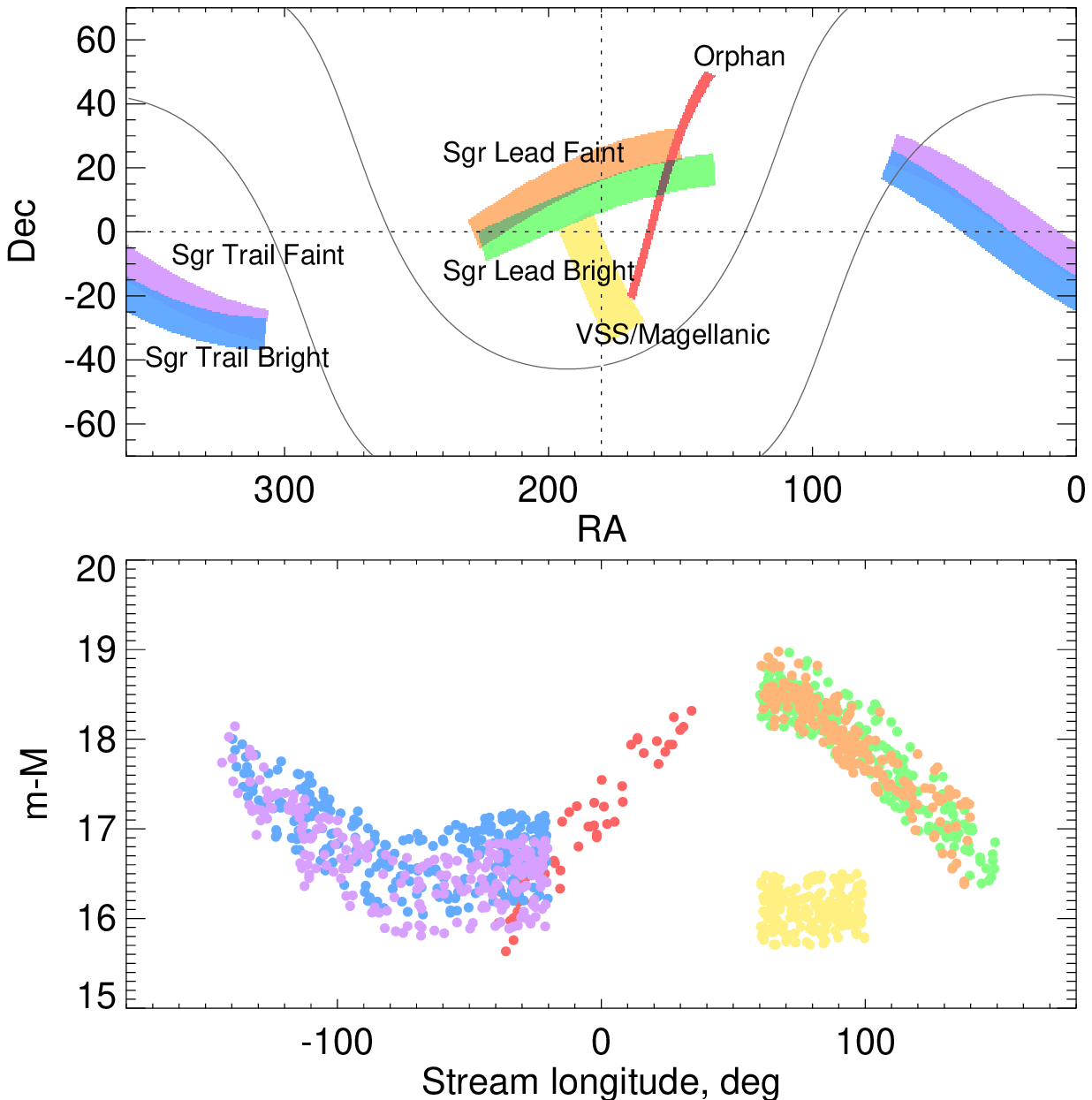}
  \includegraphics[width=0.33\textwidth]{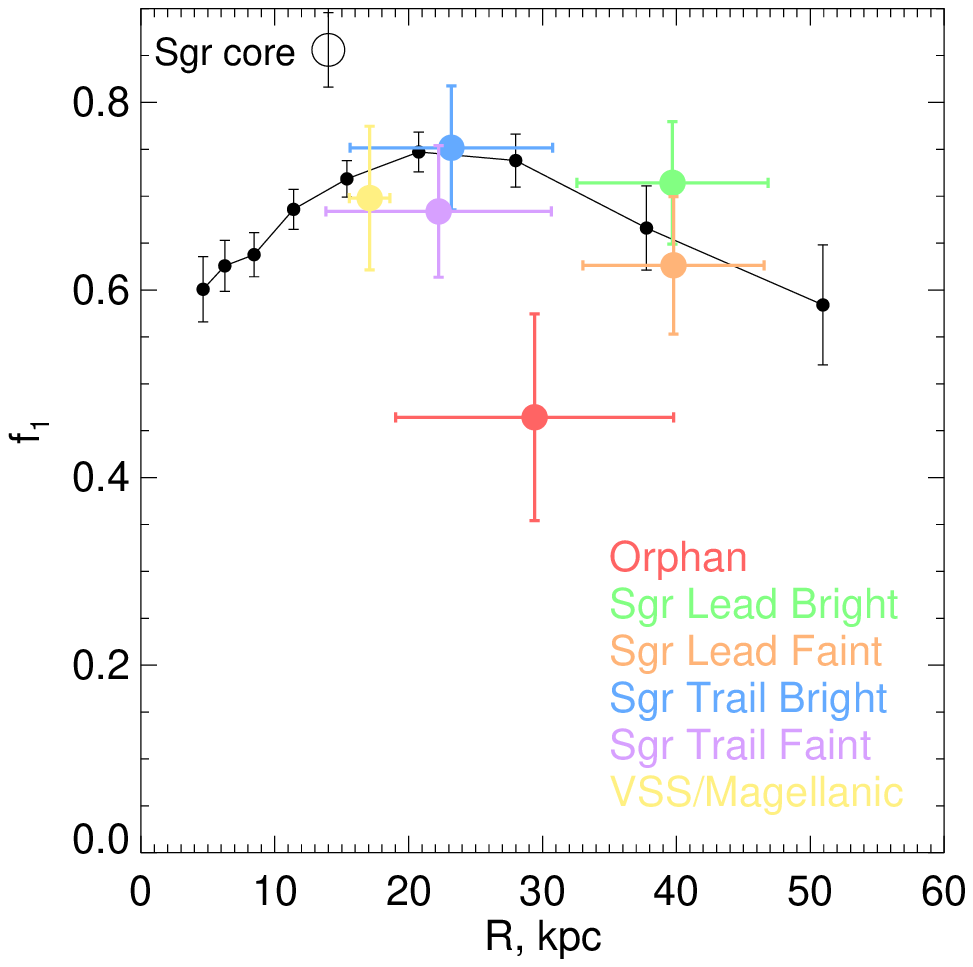}
  \includegraphics[width=0.33\textwidth]{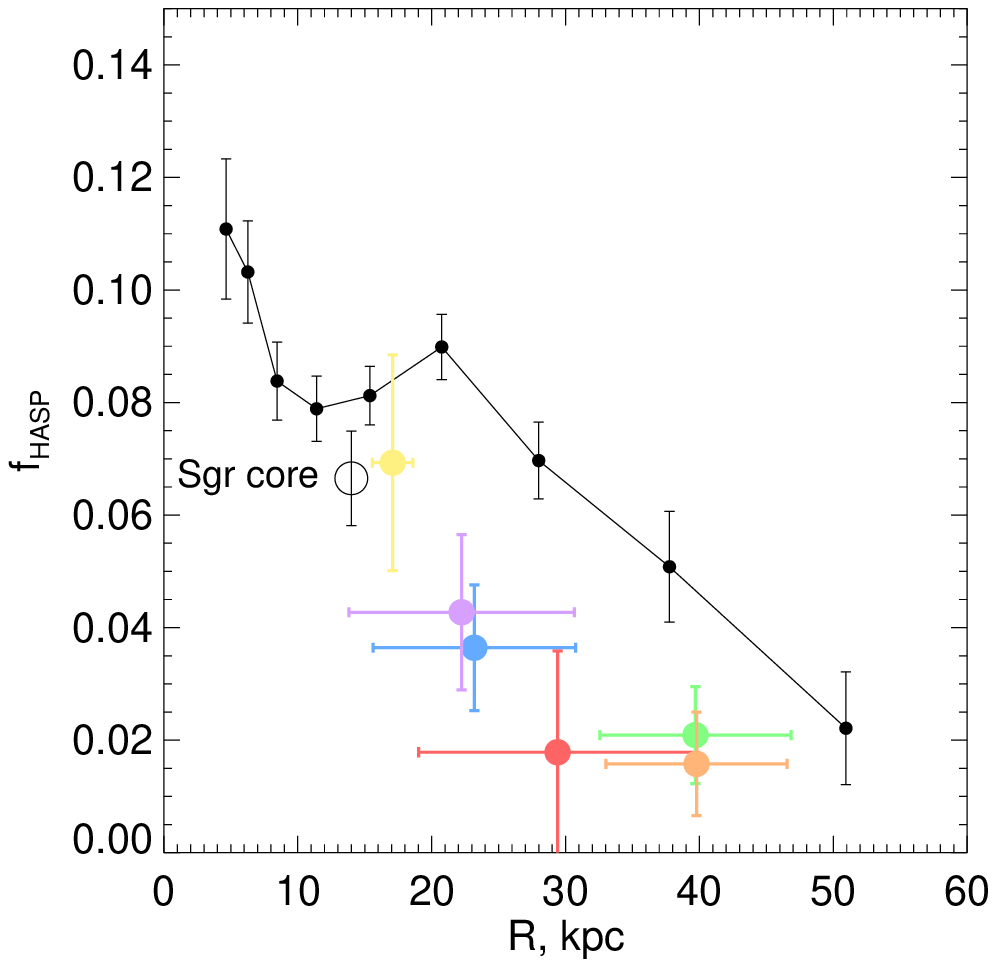}
  \caption[]{{\it Top Left:} Selection boundaries in equatorial
    coordinates for each stellar stream considered in this Paper. {\it
      Bottom Left:} Stream candidate member stars in the plane of
    stream longitude and heliocentric distance. {\it Middle:} Fraction
    $f_1$ computed for stream candidate member stars (selected as
    shown on the left) compared to the field stellar halo (black
    solid). {\it Right:} $f_{\rm HASP}$ for the selected
    streams. Empty black circle shows the measurement of the RRab
    class fraction for the Sgr dwarf core (see
    Figure~\ref{fig:dsph}).}
   \label{fig:streams}
\end{figure*}

It is instructive to compare the mixture of the field halo RRab with
that found in Galactic satellites, such as GCs and dSph
galaxies. Figure~\ref{fig:gcs} compares $f_1$ (left panel) and $f_{\rm
  HASP}$ (middle panel) curves for the field halo (as described above)
with the corresponding fractions in the GCs at the corresponding
distance. The color of the filled circle indicates the metallicity of
the cluster, with blue (red) corresponding to low (high) values of
[Fe/H]. The size of the circle is proportional to the number of RR
Lyrae in the cluster. As described in a number of previous studies ,
the GCs display a clear Oosterhoff dichotomy: $f_1$ fractions are
clustered around $\sim$0.3 and $\sim$0.8, with few examples of
intermediate values. The halo, on the other hand, occupies exactly the
range avoided by the GCs, i.e. $0.6 < f_1 < 0.75$. Unfortunately,
there are not many RR Lyrae-rich GCs in the outer halo - most have
extreme HB morphologies, either too red or too blue to produce
significant numbers of RR Lyrae. Thus, it is not possible to track the
behaviour of cluster $f_1$ with Galactocentric radius, with the
majority of the datapoints lying within 20 kpc. With regard to
$f_{\rm HASP}$, most GCs do not host as many HASP RRab stars as the
field halo at this radius. For a small number of clusters that have
more than 1 HASP star, a dichotomy similar to that of Type 1/2 is
observed. Note, however, that the cluster data agrees well with the
outer halo $F_{\rm HASP}$ fraction. Finally, the right panel of
Figure~\ref{fig:gcs} gives the track of the field halo in the space
spanned by $f_1$ and $f_{\rm HASP}$. Once again, this plot emphasizes
the lack of GCs with intermediate (0.5-0.6) values of $f_1$ and
(simultaneously) high ($\sim$0.1) values of $f_{\rm HASP}$.

As pointed out in many previous studies, the $f_1$ fraction in dSphs
as shown in Figure~\ref{fig:dsph} matches well the range of $f_1$
values in the field halo (left panel). This is the well-known
flip-side of the Oosterhoff dichotomy. In the left and right panels,
the dSph fractions are not shown at the satellites' distance but
instead are grouped around the fiducial location of 20 kpc. The color
of the symbol indicates the total stellar mass of the satellite, with
low-mass systems (like Draco and Carina) shown in purple and blue and
the largest galaxies (such as the Sgr dwarf and the Clouds) are shown
in orange and red. A clear correlation between the stellar mass and the
$f_1$ fraction is visible \citep[hints of this correlation are
  discussed in e.g.][]{Stetson2014,Fiorentino2015}. Also note that
only the most massive systems host enough Type 1 RRab stars (with the
exception of Draco) to match the peak in the field halo at
$f_1\sim$0.75 at $\sim$ 25 kpc. In terms of the HASP population
(middle panel), none of the dwarfs can attain $f_{\rm HASP}\sim 0.1$
observed in the field halo near the Galactic center and at $\sim$ 20
kpc. In fact, only three satellites, namely the Sgr, the SMC and the
LMC have enough of the HASP stars to warrant a believable
measurement. Curiously, in terms of both $f_1$ and the $f_{\rm HASP}$
values, the Sgr dwarf appears to sit around the top of the
distribution, thus indirectly confirming that the progenitor system
was one of the most massive satellites of the MW in accord with
the studies of \citet{MNO2010} and \citet{Gibbons2017}. Finally, as
the right panel of Figure~\ref{fig:dsph} illustrates, the values of
$f_1$ in the top three most massive dwarfs around the MW are too high
while, at the same time, $f_{\rm HASP}$ fractions are too low in
comparison to the field halo. These results are in good agreement with
the earlier measurements by \citet{Fiorentino2015}.

\section{Halo sub-structure}
\label{sec:streams}

Figure~\ref{fig:halomap} compares density distributions of Type 1 and
2 RR Lyrae at three different heliocentric distances. At each
distance, the halo looks remarkably different depending on the RRab
type used. The choice between Type 1 and 2 affects the appearance of
both the smooth component as well as the sub-structure. The largest
streams such as Sgr \citep[see e.g.][]{Majewski2003, FOS2006} and VSS
\citep[see][]{Duffau2006,Newberg2007,Duffau2014} are seen
predominantly in the left column, while the narrower/colder Orphan
\citep[][]{Orphan2007, Carl2006} is clearly discernible only on the
right. Moreover, splitting the halo RRab population according to the
position in the Bailey diagrams helps to clarify the sub-structure's
3D behavior. For example, the Sgr trailing stream separates cleanly
into the faint and the bright components \citep[see][]{Koposov2012}
when Type 1 objects are considered. As evidenced by the middle left
and bottom left panels of the Figure, these two branches are not only
off-set on the sky but also are located at slightly different
distances, in agreement with previous measurements
\citep[see][]{Slater2013}.

\subsection{Stellar Streams}
\label{sec:ss}

To further study the mixture of RR Lyrae stars in previously
identified halo-substructures, we select likely members of the three
stellar streams: Sgr, VSS and Orphan. Additionally, we split the Sgr
stream into four portions: the trailing and leading tails are each
divided into a bright and faint component. The RA, Dec selection
boundaries for each stream are shown in the top left panel of
Figure~\ref{fig:streams}. These regions are chosen to follow the great
circles that approximately match the average stream track on the
sky. We aid the 2D selection with a cut on heliocentric distance as
illustrated in the bottom left panel of the Figure. The resulting
measurements of $f_1$ ($f_{\rm HASP}$) fractions are given in the
middle (right) panel of the Figure. Note that, even though the
selection is performed in 3D, the samples of likely stream members do
suffer from field halo contamination. Thus the type fraction estimate
are biased towards the typical field value at the corresponding
position, in contrast with the measurements in the Galactic satellites
where the contamination is essentially negligible.

As the middle and the right panels of Figure~\ref{fig:streams}
demonstrate, even in the presence of some field contamination, a
diversity of RRab mixtures is observed in the Galactic stellar
streams. The Sgr stream and the VSS possess the highest values of
$f_1$. For Sgr, this is another indication of the high original mass
and the relatively fast enrichment history of the progenitor galaxy
\citep[see also][]{Deboer2014, Deboer2015}. Additionally, a small but
noticeable difference exists between the leading and trailing tails as
well as between their bright and faint branches. More precisely, the
RRab in the leading tail have slightly lower values of $f_1$ and
$f_{\rm HASP}$. Similarly, the faint components have a marginally
higher fraction of Type 2 RRab stars. As analysed here, the leading
debris is further away from the progenitor compared to the trailing
material, and, accordingly, is dynamically older, i.e. stripped
earlier. Thus, the difference in the RRab mixture between the leading
and trailing tails is in agreement with the metallicity gradients
observed previously along the stream \citep[see][]{Monaco2007,
  Chou2010, Hyde2015}. The difference in $f_1$ between the bright and
faint branches, if significant, may help to shed light onto the
creation of the so-called stream bifurcation
\citep[see][]{FOS2006,Koposov2012}. Currently, three models have been
put forward to explain the split in the Sgr tails
\citep[][]{Fellhauer2006, Jorge2010, Gibbons2016}. In all three
scenarios, the dynamical age of the debris in the two branches is
different, which can be exploited to link back to the stellar
populations gradients in the progenitor before the
in-fall. Superficially, the slightly lower values of $f_1$ for the
faint branch debris are consistent with the previous constraints on
the difference in metal-enrichment history of the parts of the
bifurcation \citep[][]{Koposov2012,Deboer2015}.

For the VSS, the elevated fractions $f_1$ and $f_{\rm HASP}$ may
signify much higher levels of contamination, which is not surprising
given that this is the ``fluffiest'' structure of the ones considered
here. Alternatively, this could be a clue that the VSS originated in a
massive accretion event. For example, 
\citet{Boubert2017} have recently extended the view of the stream further below
the celestial equator and pointed out that the VSS is perfectly
aligned with the Magellanic Stream (MS), thus speculating that the
structure is nothing else but the leading arm of the disrupting
MCs. Note that in all simulations of the MS, the stream is produced by
stripping the SMC rather than the LMC. Therefore, to test the
hypothesis of the VSS origin, one ought to compare the stellar
populations in the stream to those reported for the SMC. According to
the studies of \citet{Duffau2006} and \citet{Duffau2014}, the mean
metallicity of the RR Lyrae in the VSS is $\sim-1.8$, which agrees
well with the measurement of [Fe/H]$=-1.7$ for the RR Lyrae in the SMC
\citep[see][]{Haschke2012}. Intriguingly, a stream-like alignment of
several GCs coincident with the VSS was reported by
\citet{Yoon2002}.  However, unlike the stellar component of the VSS,
the GCs in the alleged stream represent a metal-poor sub-group of the
Oosterhoff II objects.

\subsection{Clustering properties}
\label{sec:cluster}

\begin{figure}
  \centering
  \includegraphics[width=0.48\textwidth]{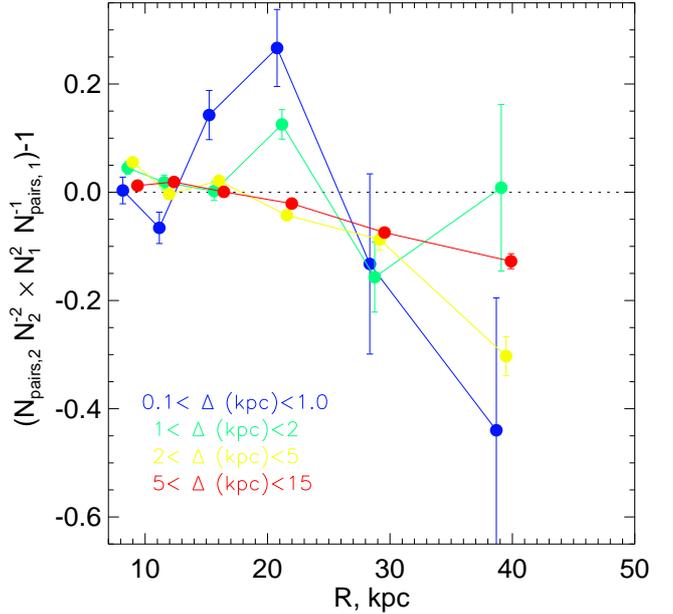}
  \caption[]{Clustering properties of the RRab stars depending on the
    type. The ratio of the number of pairs of Type 2 stars to that of
    Type 1 stars for four different 3D separations as a function of
    Galactocentric distance (in kpc). Within 15 kpc from the MW
    center, a similar amount of clustering is observed for the two
    Types. Between 15 and 30 kpc, Type 2 appears to show an excess of
    power on small scales, i.e. $<2$ kpc. This picture reverses beyond
    30 kpc, where Type 2 appears to be smoother across all scales
    compared to Type 1.}
   \label{fig:cf}
\end{figure}

Given the correlation of the fraction of Type 1 and 2 RRab stars with
the stellar mass in both the surviving dwarfs and the remnant streams
(see Section~\ref{sec:ss}), it is plausible that the clustering
properties of the smooth, i.e. well-mixed, stellar halo also depend on
the $f_1$. To this end, Figure~\ref{fig:cf} gives the ratio of the
number of pairs of RRab stars of Type 2 to Type 1 as a function of
radius for four different 3D separation scales. In other words, we
consider all combinations of two stars that fall in the same radial
bin and have 3D separation less than the given scale. RR Lyrae of both
types are affected by the CRTS selection biases. However, as we take a
ratio of the pair numbers, most of the effects due to efficiency
variations should cancel out. Note that for this calculation at each
galacto-centric radius, the number of observed pairs is scaled by the
square of the total number of stars in order to take into account the
differences in the density distributions of the RRab stars of the two
types. Furthermore, it is possible that there exist significant
differences in the large-scale on-sky distributions of the RR Lyrae in
some radial ranges. These - if indeed present - can then bias our
estimates of the relative clustering properties.

As the Figure demonstrates, three distinct regimes may exist in the
Galactic stellar halo. In the inner 10-15 kpc, the clustering
properties of Type 1 and 2 are approximately equal. This is followed
by an excess in Type 2 pair numbers out to 25-30 kpc. Beyond that, the
Type 1 appears relatively more clustered on all scales from sub-kpc to
$\sim$ 10 kpc. On the scales of 1-2 kpc and smaller, an excess of
clustering exists in the distribution of Type 2 RRab stars compared to
that of Type 1, at the Galactocentric distances between 15 and 30
kpc. The elevated number of Type 2 pairs in this radial range can be
interpreted as an increase in clumping of these RR Lyrae. However,
taking into account the behavior of $f_1$ and $f_{\rm HASP}$ fractions
(see Figure~\ref{fig:profile}), it is more likely caused by a
reduction in clustering of Type 1 stars, thus implying that this
portion of the Galactic halo is dominated by the debris from a
relatively massive progenitor system. 

\section{Discussion and Conclusions}
\label{sec:disc}

This Paper demonstrates that in the MW field halo, there exists
a strong evolution of the mixture of RRab stars as a function of
Galactocentric radius. More precisely, the fraction of Type 1
(approximately corresponding to Oosterhoff Type I) stars changes from
60$\%$ in the inner 10-15 kpc to 75$\%$ at 25 kpc, falling back to
60$\%$ beyond 40 kpc. There is a remarkable agreement between the
location of the bump in the $f_1$ fraction and the break in the radial
density of the stellar halo
\citep[see][]{Watkins2009,Deason2011,Sesar2011}. As such, our results
are in direct contradiction with those presented in \citet{Zinn2014},
but we attribute the disagreement to the differences in the size of RR
Lyrae samples used. The $f_1$ behavior can be compared to the change
in the fraction of HASP RRab variables. The $f_{\rm HASP}$ ratio also
shows an increase at around $\sim$20 kpc and an even steeper fall-off
beyond 30 kpc, where the HASP fraction drops by a factor of 5. Unlike
the Type 1 RRab profile, however, there appears to be an excess of
HASP RR Lyrae close to the Galactic centre, i.e. within 10 kpc or
so. Using simple toy models of the stellar halo, we show that the
contribution from at least three different accretion components is
required to explain the patterns in the $f_1$ and $f_{\rm HASP}$
fractions. As we elucidate in Section~\ref{sec:radial}, one plausible
interpretation of the peak in both $f_1$ and $f_{\rm HASP}$ fractions
at $\sim$25 kpc is a combination of i) different flattening of the RR
Lyrae with shorter period and ii) a sharp truncation of the distribution
of the RR Lyrae with short periods. Both of these conditions may be
accommodated in a scenario with an early accretion of a single massive
system \citep[see][]{Deason2013}.

\begin{figure*}
  \centering
  \includegraphics[width=0.98\textwidth]{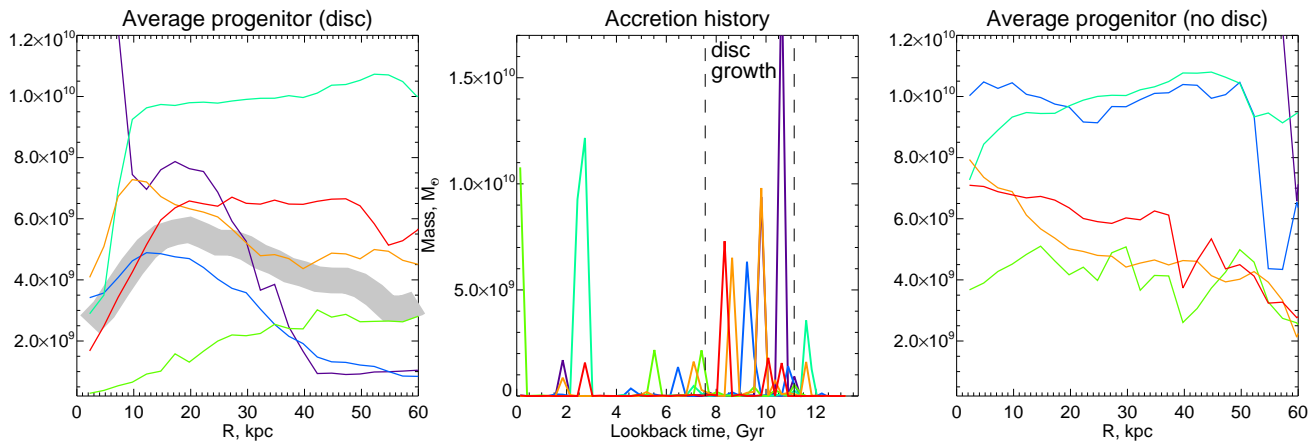}
  \caption[]{{\it Left:} Mean progenitor mass at the given
    Galactocentric distance in six simulated MW stellar halos. The thick
    grey band represents the average between the blue and the red
    curves. Color is used to distinguish between individual halos and
    link to the accretion history shown in the middle panel. {\it
      Middle:} Accretion histories of the simulated stellar
    halos. {\it Right:} Same as left panel, but for stellar halos
    simulated without a baryonic disc. Note that the disc grown
    between redshifts 3 and 1 affects the debris distribution in the
    stellar halo and can sometimes turn a rather flat progenitor mass
    profile (as shown in the right panel) into a ``bumpy'' one (as
    shown in the left panel).}
   \label{fig:sims}
\end{figure*}

To help calibrate the $f_1$ and $f_{\rm HASP}$ estimates, we also
analyze the mixture of RRab stars in the surviving MW satellites, such
as GCs and dwarfs spheroidals (see Figures~\ref{fig:gcs}
and \ref{fig:dsph}). As noted previously by many authors, the GCs
show a clear dichotomy, with some preferring low values of $f_1$ and
some high. The dSphs, on the other hand, tend to have intermediate
values of $f_1$. While many of the GCs lie either too low or too high
compared to the $f_1$ in the field halo, dSphs appear to match the
range of the observed halo Type 1 fraction much better. We register a
correlation between the dwarf's stellar mass and $f_1$, with the three
most massive satellites, namely Sgr, the SMC and the LMC, reaching
$f_1\sim80\%$, slightly above the halo peak of $f_1\sim 75\%$. Even
though there exist objects amongst both the GCs and the dSphs with
sufficiently high $f_1$ values, neither of the two satellite classes
contains fractions of HASP RR Lyrae similar to the halo. The bulk of
the GCs and the dSphs (including the most massive ones) contain a
factor of two lower numbers of HASP variables. There is a small number
of GCs with fractions $f_{\rm HASP}$ a factor of 1.5 higher compared
to the peak of the field halo. However, as the right panel of
Figure~\ref{fig:gcs} illustrates, the exact combination of
intermediate $f_1$ and elevated $f_{\rm HASP}$ is not realised in any
of the surviving satellites, be it a GC or a dSph.

Stepping aside from studying the average properties of the RR Lyrae
mixture in the field, maps of the RRab density shown in
Figure~\ref{fig:halomap} reveal striking differences in the properties
of the stellar halo depending on the Type of the pulsator used. We
choose three well-known halo sub-structures whose 3D properties have
been mapped out previously, namely the Sgr Stream, the Virgo Stellar
Stream and the Orphan Stream. We detect hints of evolution of $f_1$
along the Sgr tails, as well as subtle $f_1$ differences between the
bright and the faint branches of the stream. We hypothesize that our
measurements are consistent with chemical abundances gradients in the
progenitor. If confirmed, these ought to help elucidate the models of
the Sgr dwarf disruption and the genesis of the stream
bifurcation. Surprisingly, we find that the VSS possesses high values
of both $f_1$ and $f_{\rm HASP}$, which would imply its origin in a
massive galaxy, perhaps similar to the SMC. If not a result of a field
halo contamination our measurement lends support to the recent
discovery of \citet{Boubert2017}, who point out a close connection
between the extended view of the VSS they uncover and the Magellanic
Clouds. Finally, the Orphan Stream contains the lowest fraction of the
Type 1 RR Lyrae, $f_1\sim 0.5$, which, given a likely non-zero
contamination is only an upper bound on its true $f_1$. Our
measurement of $f_1$ for the Orphan Stream has two important
consequences. First, it hints at the possibility of extending the
$f_1$-mass relationship to lower mass objects, such as ultra-faint
dwarfs. Second, it shows that the search for low-mass sub-structure in the
halo can be significantly improved by using a particular sub-set of
RRab stars, more precisely those with lower values of $f_1$.

What is the most straightforward interpretation of the radial
evolution of $f_1$ fraction as shown in Figure~\ref{fig:profile}?
Given the correlation between the stellar mass and the $f_1$ fraction
exhibited by the dSphs currently on orbit around the MW, the Galactic
$f_1$ profile may simply be a reflection of the change in the
fractional contribution of the halo progenitors of different
masses. In other words, at distances below 15 kpc and beyond 40 kpc
where $f_1$ is at its lowest, an enhanced contribution from lower mass
systems is expected. Between 15 and 30 kpc, where the peaks in both
$f_1$ and $f_{\rm HASP}$ are observed, the stellar halo is dominated
by the debris from a massive progenitor --- a hypothesis similar to the
halo break theory put forward by \citet{Deason2013}. To test this
conjecture, we analyse a suite of Cosmological zoom-in simulations of
MW halo formation. The details of the simulations can be found
in \citet{Jethwa2016}. In essence, the six simulated stellar halos
considered here sample a range of MW masses and capture
some of the effects induced by the baryons present in the real Galaxy,
i.e. the action of the disc. The baryonic disc is implemented
parametrically and is grown adiabatically between redshifts 3 and
1. To create the stellar halo, the $3\%$ most bound dark matter
particles in each in-falling sub-halo are tagged as stars. This is done by determining
when each subhalo reaches its peak mass and then assigning it a stellar mass of
\begin{equation}
 M_* = 3.4\times10^6 M_\odot \left( \frac{M_{\rm peak}}{10^9 M_\odot} \right)^{1.7} ,
 \end{equation}
similar to the approach in \cite{delucia_helmi08,bailin+14}. In order to only sample the 
stellar halo, stars within 2 tidal radii of the remaining subhaloes at $z=0$ are excluded. 

The left panel of Figure~\ref{fig:sims} presents the average progenitor
stellar mass as a function of Galactocentric radius in the six
simulated halos whose accretion histories are displayed in the middle
panel. All (but one) halos display a drop in the average progenitor
mass at small distances from the halo's center. The fake MWs
with extended or recent accretion (shades of green) display a flat
mass profile across the range of distances considered here,
i.e. between 20 and 70 kpc. Those halos whose accretion history peaked
early, i.e. those colored purple, blue, orange and red all show a drop
in the average progenitor mass beyond 30-60 kpc. Clearly, the exact
shape of the mean progenitor mass profile is very sensitive to the
details of the host accretion history, which cannot be exhaustively
sampled with a small number of simulations presented here. For
illustration purposes, we also show a combination of the two mean mass
profiles (for the halos shown in red and blue) as a thick grey
line. As demonstrated in the Figure, MWs with an accretion
history peaking between 8 and 10 Gyr ago can indeed possess a stellar
halo with a characteristic bump in the mean progenitor mass profile,
i.e. very similar to the one implied by the RRab studies presented
here. Curiously, as shown in the right panel of Figure~\ref{fig:sims},
the corresponding zoom-in simulations without a disc do not show any
of the trends discussed above. Regardless of the accretion history,
between 5 and 50 kpc, the mean progenitor mass profiles appear rather
flat. We therefore conclude that the changes in the mean mass profile
as seen in the left panel of the Figure are put in place by the action
of the disc. As the disc grows from $\sim$11 to $\sim$8 Gyr ago, it
helps to migrate the stellar debris accumulated so far from the outer
parts of the halo into the inner regions. Around the Galactic enter,
this ``debris sorting'' induced by the disc growth creates an excess
of the material accreted the earliest (11-13 Gyr) --- and thus mostly
contributed by the lower mass objects (as bigger dwarfs require some
time to grow before falling into the MW). The stars stripped from the
larger systems later on (8-11 Gyr) are moved closer to the Galactic
center but can not be packed as tightly hence resulting in
over-density at intermediate radii (20-30 kpc).  Therefore, the three
stellar halo components required by the RR Lyrae mixture changes could
be i) the debris from low-mass objects accreted early (10-13 Gyr) and
contributing to the inner region of the halo today ($<$15 kpc), ii)
material from a massive system merging with the MW between 8 and 11
Gyr ago and dominating the halo at the distances between 10 and 30 kpc
and iii) the stars from the rest of the unlucky Galactic satellites
fallen into the MW during its lifetime.

To summarize, the changes in the RRab mixture with radius can be
explained with the evolution of the relative contribution of
progenitors of different masses. Given the behavior of $f_1$, the
field halo at low and high Galactocentric distances may be dominated
by debris from systems not unlike a typical dSph observed around the
MW today. The intermediate distances in the Galaxy appear to have an
increased contribution from the most massive systems, perhaps as
massive as Sgr/SMC/LMC. However, the exact nature of such a ``giant
dwarf'' progenitor is unclear as the halo's $f_{\rm HASP}$ fraction at
these distances is almost twice that of the most massive surviving
dwarf (LMC). Of course, such an ancient progenitor system could not
necessarily be a replica of the LMC. Note that the LMC had had a
prolonged period of inactivity, restarting star-formation in earnest
only 5 Gyrs ago \citep[see][]{Harris2009}. The hypothesised halo
progenitor, on the other hand, evolved much closer to the MW, and thus
likely had a different --- probably, faster --- enrichment history. If
a large portion of Type 1 RR Lyrae stars around the 15-30 kpc has
indeed been contributed via an early merger with a massive system, it
may be possible to pick up tracers of this accretion event in the
kinematics of stars in the future Gaia data releases. Note that many
Oosterhoff I GCs tend to prefer retrograde orbits according to
\citet{vdb1993_oo,vdb1993_oo2} who suggested the merger of a massive
ancestral object(s) on a retrograde orbit as the origin of the MW's
OoI component. With regards to the short period RRab stars, it appears
rather plausible that the sharp rise in $f_{\rm HASP}$ near the
Galactic center is due to the bulge, in agreement with the earlier
studies of \citet{Kunder2009}. Finally, our hypothesis of the strong
evolution in the mass of the ``typical'' contributing progenitor as a
function of Galactocentric radius appears to be supported by the
tentative results of the clustering analysis of the RRab stars
presented in Section~\ref{sec:cluster} and by the comparison to the
Cosmological zoom-in simulations of the MW halo formation in the
presence of a baryonic disc.

The ideas discussed here are complementary to the analysis of the
stellar population mixture as encoded in colour-magnitude space
\citep[see e.g.][]{deJong2010} and has many similarities with other
studies of variations in the ratio of distinct halo tracers
\citep[see][]{Bell2010,Deason2015,PW2015}. Given the diversity of the
behaviour of the simple test statistics proposed here, such as $f_1$
and $f_{\rm HASP}$ fractions, we surmise that the distribution of RR
Lyrae in the period-amplitude space appears to be unique enough to be
used as a progenitor fingerprint. In particular, it may be possible to
use it to tease out signatures of the accretion of low-mass
satellites. There is also an added value in being able to leverage the
unique properties of the RR Lyrae as standard candles to disentangle
the hotchpotch of the stellar halo in 3D. All in all, in the view of
the imminent Gaia DR2 release, unmixing the Galactic halo with RR
Lyrae tagging should become feasible in the near future.

\section*{Acknowledgments}

The authors wish to thank Kathryn Johnston and Douglas Boubert for
stimulating discussions that helped to improve this paper. VB is
grateful to Nat\`alia Mora-Sitj\`a for the careful proof-reading of
the manuscript.

The research leading to these results has received funding from the
European Research Council under the European Union's Seventh Framework
Programme (FP/2007-2013) / ERC Grant Agreement n. 308024.
A.D. is supported by a Royal Society University Research Fellowship. 
A.D. also acknowledges support from the STFC grant ST/P000451/1. NWE
thanks the Center for Computational Astrophysics for hospitality
during a working visit.

The authors gratefully acknowledge support by CONICYT/RCUK's PCI
program through grant DPI20140066 (``Newton Funds''). M.C. is
additionally supported by the Ministry for the Economy, Development,
and Tourism's Millennium Science Initiative through grant IC\,120009,
awarded to the Millennium Institute of Astrophysics (MAS); by Proyecto
Basal PFB-06/2007; and by FONDECYT grant \#1171273.

\bibliography{references}

\label{lastpage}

\end{document}